\documentclass[prd,nofootinbib,aps,superscriptaddress,tightenlines,preprintnumbers]{revtex4}

\usepackage{latexsym,amsmath,amssymb,graphicx,hyperref}

\def\a{\alpha}		\def\b{\beta}

\def\m{\mu}		\def\n{\nu}			\def\o{\omega}
\def\p{\pi}					\def\r{\rho}
\def\s{\sigma}

\def\aether{{\ae}ther}
\def\Cerenkov{\v Cerenkov}

\def\kv{\vec{k}}

\def\be{\begin{equation}}
\def\ee{\end{equation}}
\begin{document}
\preprint{CALT-68-2713}
\title{Sigma-Model {\AE}ther}
\author{Sean M. Carroll, Timothy R. Dulaney, Moira I. Gresham, and Heywood Tam}
\email[]{seancarroll@gmail.com}
\email[]{dulaney@theory.caltech.edu}
\email[]{moira@theory.caltech.edu}
\email[]{tam@theory.caltech.edu}
\affiliation{California Institute of Technology,~Pasadena, CA 91125, USA}
\date{\today}

\begin{abstract}
Theories of low-energy Lorentz violation by a fixed-norm ``\aether'' vector field with two-derivative kinetic terms have a globally bounded Hamiltonian and are perturbatively stable only if the vector is timelike and the kinetic term in the action
takes the form of a sigma model.  Here we investigate the phenomenological
properties of this theory.   We first consider the propagation of modes in the
presence of gravity, and show that there is a unique choice of curvature coupling 
that leads to a theory without superluminal modes.  Experimental constraints on 
this theory come from a number of sources, and we examine bounds in a two--dimensional
parameter space.  We then consider the cosmological evolution of the \aether,
arguing that the vector will naturally evolve to be orthogonal to constant-density
hypersurfaces in a Friedmann-Robertson-Walker cosmology.  Finally, we 
examine cosmological evolution in the presence of an extra compact dimension of
space, concluding that a vector can maintain a constant projection along the
extra dimension in an expanding universe only when the expansion is exponential.
 \end{abstract}
\preprint{}
\maketitle

\section{Introduction} 

Models of fixed-norm vector fields, sometimes called ``{\ae}ther'' theories, serve a useful
purpose as a phenomenological framework in which to investigate violations of Lorentz
invariance at low energies \cite{Kostelecky:1989jw, Colladay:1998fq, Jacobson:2000xp, Eling:2003rd, Carroll:2004ai, Jacobson:2004ts, Lim:2004js}.  For a recent review, see \cite{Jacobson:2008aj}.  In a companion paper \cite{instabilities}, we argue that almost all such models are plagued by instabilities.   For related work on stability in  \aether\ theories, see \cite{Kostelecky:2000mm, Kostelecky:2001xz, Clayton:2001vy, Elliott:2005va, Dulaney:2008ph, Bluhm:2008yt, Himmetoglu:2008zp, Jimenez:2008sq}.  

There is one version of the \aether\  theory that is stable under small perturbations and in which the Hamiltonian is globally bounded when only two-derivative terms are included in the action.
This model is defined by a kinetic Lagrange density of the form
\be
  {\cal L}^{\text{kinetic}}_\sigma = -\frac{1}{2}(\nabla_\mu A_\nu)(\nabla^\mu A^\nu)\,,
  \label{sigmakinetic}
\ee
where $A_\mu$ is a dynamical timelike four-vector \aether\ field.  (The spacelike version
has an unbounded Hamiltonian, and is unstable.)
We refer to the theory defined by this action as ``sigma-model \aether,'' due to its
resemblance to a theory of scalar fields propagating on a fixed manifold with an
internal metric, familiar from studies of spontaneous symmetry breaking.    
The \aether\ theory is not identical to
such a sigma model---in particular in curved space where covariant derivatives act
on the vector---but the nomenclature is convenient.

Even though this theory is stable, it has an important drawback.  It is conventional in
\aether\ models to give the vector field an expectation value by means of a Lagrange
multiplier, which enforces the fixed-norm constraint
\be
  A_\mu A^\mu = - m^2\,.
  \label{norm}
\ee
We take $m^2$ to be positive and use a metric signature $(-+++)$, so that this
defines a timelike vector field.  Despite the convenience of this formulation, it seems
likely that a more complete version of the theory would arise as  a limit of a theory in which
the expectation value is fixed by minimizing a smooth potential of the form $V(A_\mu) = \xi
(A_\mu A^\mu + m^2)^2$.  As we showed in \cite{instabilities}, any 
such theory would be plagued by ghosts and tachyons.  As far as we can tell, therefore,
the sigma-model \aether\ theory cannot be derived from models with a smooth potential.

Nevertheless, as it is the only example of a Lorentz-violating \aether\ theory that we are sure
is globally well-behaved, examining the dynamics and experimental constraints on this model is worthwhile.  We undertake such an investigation in this paper.  

First we examine the degrees of freedom in 
this theory, taking into account the mixing with the gravitational field.  There are three different massless modes, of spins 0, 1, and 2 in the \aether\ rest frame.\footnote{The lack of rotational symmetry in frames other than the \aether\ rest frame make classification of modes by spin in such frames impossible. But the \aether\  rest frame has rotational symmetry, which allows for the spin classification with respect to this frame.}  Demanding that none of the modes propagate faster
than light fixes a unique value for the coupling of the vector field to the Ricci tensor.  We use experimental constraints on the preferred frame parameters $\alpha_{1,2}$ in the Parameterized Post-Newtonian (PPN) expansion to limit the magnitude of the vacuum expectation value, $m$. 
The spin-2 mode can propagate subluminally for some values of the vector field/Ricci tensor coupling; in such cases, very tight restrictions on the vacuum expectation value, $m$, due to limits from vacuum \Cerenkov\ radiation of gravitons come into play.  

Finally, we consider the cosmological evolution of the vector field in two different backgrounds.  We study the evolution of the timelike vector field in a flat Friedmann-Robertson-Walker (FRW) universe and find that the vector field tends to align to be orthogonal to constant density hypersurfaces.   In a background consisting of a timelike dimension, three expanding spatial dimensions, and one compact (non-expanding) extra spatial dimension, we find that the  vector field can evolve to have a non-zero projection in the direction of the compact extra dimension if the large dimensions are de~Sitter-like.  We take this as evidence that a timelike vector field with the Lagrangian that satisfies the aforementioned theoretical and experimental constraints would not lead to any significant departure from isotropy. 

\section{Excitations in the Presence of Gravity}

We would like to understand the experimental constraints on, and cosmological evolution
of, the sigma-model \aether\ theory.  For both of these questions, it is important to 
consider the effects of gravity.  But whereas the flat-space model with a kinetic
Lagrangian of the form (\ref{sigmakinetic}) is unique, in curved space there is
the possibility of an explicit coupling to curvature.  The full action we consider is
\begin{equation}
	\label{curved space action}
  S = \int d^4x \sqrt{-g} \left[\frac{1}{16\pi G}R - {1\over 2}( \nabla_\m A_\n)(\nabla^\m A^\n) 
  + {\alpha \over 2} R_{\m\n}A^\m A^\n 
  + {\lambda \over 2} (A_\m A^\m + m^2)\right]\,.
\end{equation}
Here, $\lambda$ is the Lagrange multiplier that enforces the fixed-norm constraint
(\ref{norm}), $\alpha$ is a dimensionless coupling, $R_{\mu\nu}$ is the Ricci tensor
and $R$ is the curvature scalar.  Note that, given the fixed-norm constraint, there are
no other scalar operators that could be formed solely from $A_\mu$ and the Riemann
tensor $R^\rho{}_{\sigma\mu\nu}$.
By integrating by parts and using $R_{\m \n}A^\m A^\n = A^\n[\nabla_\m,\nabla_\n]A^\m$,
this curvature coupling could equivalently be written purely in terms of covariant
derivatives of $A_\mu$; the form (\ref{curved space action}) has the advantage of
emphasizing that the new term has no effects in flat spacetime.

In \cite{instabilities} we showed that the sigma-model \aether\ theory was stable in the presence
of small perturbations in flat spacetime; the possibility of mixing with gravitons implies that
we should check once more in curved spacetime.
The equations of motion for the vector field are
\begin{equation}
-\nabla_\m \nabla^\m A^\n = \lambda A^\n + \a R^{\m \n}A_\m,
\end{equation}
along with the fixed norm constraint from the equation of motion for $\lambda$.
Assuming the fixed norm constraint, the equations of
motion can be written in the form
\begin{equation}
\left(g^{\s \n} + \frac{1}{m^2}A^\s A^\n\right)(\nabla_\r \nabla^\r A_\s + \a R_{\r \s}A^\r) = 0.
\end{equation}
The tensor $(g^{\s \n} + A_\rho A_\nu /m^2)$ acts to take what would be the equation
of motion in the absence of the constraint, and project it into the hyperplane orthogonal
to $A_\mu$.

The Einstein-\aether\ system has a total of five degrees of freedom, all of which 
propagate as massless fields:  one spin-2 graviton, one spin-1 excitation and one spin-0 excitation.  Each of these dispersion relations can be written (in the short-wavelength limit) in frame-invariant notation as
\begin{equation}
	\label{disp}
k_\m k^\m  = \left(\frac{1-v^2}{v^2}\right)\left(\frac{\bar{A}_\m k^\m}{m} \right)^2,
\end{equation}
where $v$ is the phase velocity in the {\ae}ther rest frame.  The squared phase velocities of the gravity-{\ae}ther modes  are \cite{Jacobson:2004ts}
\begin{equation}
	\label{spin-2}
v^2_2 = {1 \over 1 - 8 \p G m^2(1 + \a)} \approx 1 + 8 \p G m^2 (1 + \a) \qquad (\text{spin-2}) 
\end{equation}
\begin{equation}
	\label{spin-1}
v^2_1 = {2 - 8 \p G m^2 (1 + \a)(1 - \a) \over 2\left(1 - 8 \p G m^2 (1 + \a) \right)} \approx 1 + 4 \p G m^2(1 + \a)^2 \qquad  (\text{spin-1})
\end{equation}
\begin{equation}
	\label{spin-0}
v^2_0 = {2 - 8 \p G m^2 \over \left(1 - 8 \p G m^2(1 + \a)\right)\left( 2 + 8 \p G m^2 (1 - 2 \a) \right)} \approx 1 + 16 \p G m^2 \a \qquad (\text{spin-0})
\end{equation}
where $G$ is the gravitational constant appearing in Einstein's action.  The approximate equalities hold assuming $8 \p G m^2 \ll 1$.\footnote{The relationship between the parameters in Eq.~\eqref {curved space action} ($\a$, $m^2$) and those in Ref.~\cite{Jacobson:2004ts} ($c_1, c_2, c_3, c_4$) is: 
\begin{equation}
c_1 = 8 \p G m^2, \qquad -c_2 = c_3 = \a 8 \p G m^2, \qquad c_4 = 0.
\label{translation}
\end{equation}
}

\begin{figure}
\centering
\includegraphics[width=0.6\textwidth]{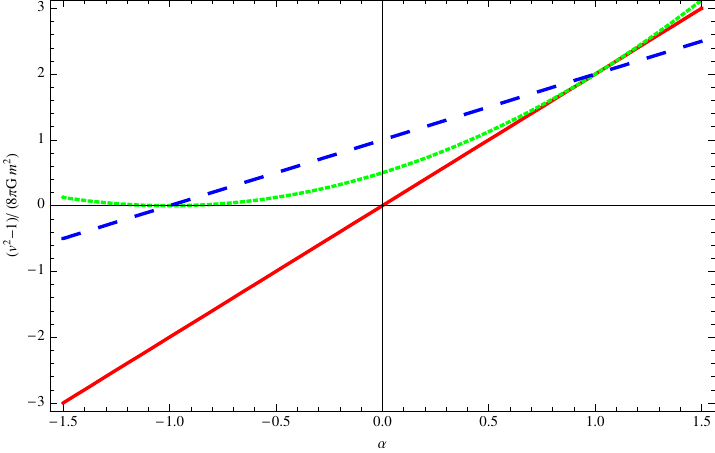}
\caption{{\AE}ther rest frame mode phase velocities squared, $v^2$, minus the speed of light in units of $8 \p G m^2$ as a function of $\a$. The solid (red) line corresponds to spin-0, the small dashing (green) to spin-1, and the large dashing (blue) to spin-2.  Only for $\alpha=-1$ do none of the modes propagate
faster than light ($v^2 - 1 > 0$).}
\label {mode speed plots}
\end{figure}

These squared mode phase velocities minus the squared speed of light are plotted in Fig.~\ref{mode speed plots} as a function of $\a$. It is clear that the only value of $\alpha$ for which none of
the modes propagate superluminally ($v^2 > 1$) is
\be
  \alpha = -1\,.
\ee
We therefore have a \emph{unique} version of a Lorentz-violating \aether\ theory 
for which the Hamiltonian is bounded below (in flat space) and that is free of superluminal
modes when coupled to gravity:  the sigma-model
kinetic term with an expectation value fixed by a Lagrange-multiplier constraint and a 
coupling to curvature of the form in (\ref{curved space action}) with $\a=-1$.  In what
follows, we will generally allow $\a$ to remain as a free parameter when considering
experimental limits, keeping in mind that models with $\a\neq -1$ are plagued by
superluminal modes.  We will find that the experimental limits on $m$ are actually weakest
when $\a = -1$.

Before moving on, however, we should note that the existence of superluminal phase
velocities does not constitute \emph{prima facie} evidence that the theory is ill-behaved.
There are two reasons for suspecting that superluminal propagation is bad.
First, in \cite{instabilities}, we showed that such models
were associated with perturbative instabilities:  there is always a frame in which
small perturbations grow exponentially with time.  Second, acausal propagation around
a closed loop in spacetime could potentially occur if the background \aether\ field were not constant
through space \cite{Carroll:2004ai, Elliott:2005va}. 
But in the presence of gravity, these arguments are not decisive.  There now exists
a scale beyond which we expect the theory to break down: namely, length scales on the order of $M_{pl}^{-1}$. Perhaps there is some length scale involved in boosting to a frame where the instability is apparent (or, equivalently, in approaching a trajectory that is a closed timelike curve) that is order $M_{pl}^{-1}$.

Again, in a background flat spacetime with a background timelike \aether\ field $\bar{A}_\mu = $~constant, the
dispersion relations have the generic form
\begin{equation}\label{Tdispersion}
(v^{-2} - 1) (t^\m k_\m)^2 = k_\m k^\m,
\end{equation} 
where $t_\m = \bar{A}_\m/m$ characterizes the 4-velocity of the preferred rest frame. 
The velocity $v^2$ is given by Eqs.~\eqref{spin-2}-\eqref{spin-0}. In a boosted frame, where $t^\m = (-\cosh \eta, \sinh \eta \, \hat{n})$, the frequency is given by 
\begin{equation}
{\omega \over |\kv|} = {-(1 - v^{-2})\sinh \eta \cosh \eta (\hat{k} \cdot \hat{n}) \pm \sqrt{1 - (1 - v^{-2})(\cosh^2\eta - \sinh^2 \eta (\hat{n} \cdot \hat{k})^2)} \over 1 - (1 - v^{-2}) \cosh^2 \eta}.
\end{equation}
Let us parameterize the boost in the standard way as
\begin{equation}
\cosh^2 \eta = {1 \over 1 - \b^2}\ , \qquad 0 \leq \b^2 < 1.
\end{equation} 
Then 
\begin{align}
{\omega \over |\kv|} 	
				= {-(1 - v^{-2}) \b (\hat{k} \cdot \hat{n}) \pm \sqrt{1 - \b^2} \sqrt{v^{-2} - \b^2 + \b^2(1 - v^{-2}) (\hat{n} \cdot \hat{k})^2} \over  v^{-2} - \b^2 }.
\end{align}
There is a pole in the frequency at $\b^2 = v^{-2}$. The pole is physical if $v > 1$ and, (in the limit as $\hat{n}\cdot \hat{k} \rightarrow 0$) as $\b$ passes through the pole ($\b^2 \rightarrow \b^2 > v^{-2}$), the frequency acquires a nonzero imaginary part, which corresponds to growing mode amplitudes. (The frequency becomes imaginary at some $\beta^2 < 1$ as long as $\hat{n}\cdot\hat{k} \neq 1$.) The time scale on which the mode grows is set by $1 / Im(\o)$. In frames with a boost factor greater than the inverse rest-frame mode speed, $\b > v^{-1}$, the time scale on which mode amplitudes grow is maximal for modes with wave vectors perpendicular to the boost direction ($\hat{n} \cdot \hat{k} = 0$) and is given by
\begin{equation}
T_{MAX}(\b) = {1 \over | Im(\omega) | }  = | \vec{k}|^{-1} { \sqrt{\b^2 - v^{-2}} \over \sqrt{1 - \b^2} } \qquad  \text{when} \qquad v^2 > 1.
\end{equation}  

We generically expect the linearized gravity analysis that led to the propagation speeds in Eqs.~\eqref{disp}-\eqref{spin-0} to be valid for wave vectors that are much greater in magnitude than the energy scale set by other energy density in the space-time---generally, the Hubble scale, $H$. Thus the analysis makes sense for $|\vec{k}|^{-1} \ll H^{-1}$ and (as long as $1 - \b^2$ is not infinitesimal) there will be instabilities on time scales less than the inverse Hubble scale and (unless $\b^2 - v^{-2}$ is infinitesimal) greater than $M_{Pl}^{-1}$.  

Thus, not only could superluminal propagation speeds lead to closed timelike curves and violations of causality, but the existence of instabilities on an unremarkable range of less-than-Hubble-radius time scales in boosted frames indicates that such superluminal propagation speeds lead to instabilities.  If $v > 1$, it appears as if instabilities can be accessed without crossing some scale threshold beyond which we'd expect the model to break down.

\section{Experimental Constraints}

We now apply existing experimental limits to the sigma-model \aether\ theory,
keeping for the moment $\a$ as well as $m^2$ as free parameters.  Direct coupling of the \aether\ field to Standard Model fields fits into the framework of the ``Lorentz-violating extension'' of the Standard Model considered in Ref.~\cite{Colladay:1998fq}.  Such couplings are very tightly constrained  by various experiments (for a discussion of experimental constraints, see Ref.~\cite{Mattingly:2005re}).  The relevant limit from gravitational \Cerenkov\ radiation in \cite{Elliott:2005va} translates to\footnote{Ref.~\cite{Elliott:2005va} uses the same parameters as in \cite{Jacobson:2004ts, Jacobson:2008aj}, thus the translation between our parameters and the parameters used in \cite{Jacobson:2004ts, Jacobson:2008aj, Elliott:2005va} is as stated in (\ref{translation}).}
\begin{align}
	\label{cherenkov constraints}
-8 \p G m^2(1 + \a)  < 1 \times 10^{-15}.
\end{align}

\begin{figure}
\centering
\includegraphics[width=0.45\textwidth]{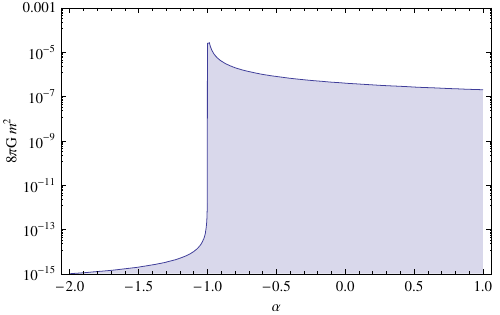}
\includegraphics[width=0.45\textwidth]{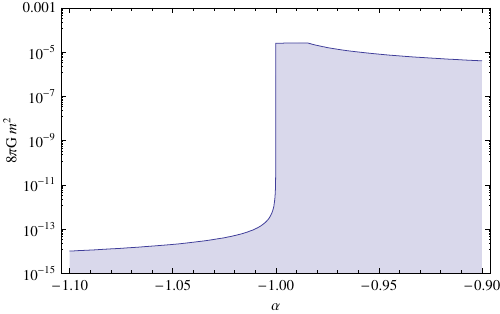}
\caption{Parameter space allowed (shaded region) by constraints from \Cerenkov\ radiation and PPN. The strongest constraint in the $\a < -1$ region is from Eq.~\eqref{cherenkov constraints}, and for most of the $\a > -1$ region the strongest constraint is from the second inequality in Eq.~\eqref{ppn constraint}. The plot on the right is a blow-up of the small range of $\a$ for which the first constraint in Eq.~\eqref{ppn constraint} is strongest---when $\alpha = -1$ to within a couple of parts in one hundred. }\label{constraints fig}
\end{figure}

Limits on PPN parameters give the strongest constraints on $\a$ and $m^2$ when $\a \approx -1$ (since the constraint in Eq.~\eqref{cherenkov constraints} is automatically satisfied).  The preferred frame parameters must satisfy $|\alpha_1| < 10^{-4}$ and $|\alpha_2| < 10^{-7}$ \cite{Will:2005va}.  We have the limits  \cite{Jacobson:2008aj}
\begin{align}
|\alpha_1| \approx |4\a^2(8\pi G_N m^2)| < 10^{-4}  ~~\text{and} ~~ |\alpha_2| \approx |(\a+1)(8\pi G_N m^2)| < 10^{-7},
\label{ppn constraint}
\end{align}
where $G_N$ is the gravitational constant as measured in our solar system or table-top experiments.  This gravitational constant is related to the parameter in the action $G$ by \cite{Carroll:2004ai}
\begin{equation}
G_N = {G \over 1 - 4\pi G m^2}.
\end{equation}
If we require that all modes have phase speeds $v$ that satisfy $v^2 \le 1$, then we must have $\a = -1$ and 
\begin{equation}
8\pi G_N m^2 < 10^{-4} \qquad (\a = -1).
\end{equation}  All relevant constraints (allowing modes to have larger than unity phase velocities) are summarized in Fig.~\ref{constraints fig}.  Constraints from Big Bang Nucleosynthesis \cite{Carroll:2004ai} are significantly weaker than the PPN and \Cerenkov\ constraint above.

\section{Cosmological Evolution}

We now turn to the evolution of the sigma-model \aether\ field in a cosmological
background.   It is usually assumed in the literature that the \aether\ preferred frame coincides
with the cosmological rest frame---{\it i.e.}, that in Robertson-Walker coordinates, a timelike
\aether\ field has zero spatial components, or a spacelike \aether\ field has zero time component.
Under this assumption, there has been some analysis of cosmological evolution in the presence of \aether\ fields \cite{Ackerman:2007nb, Dulaney:2008ph, Dulaney:2008bp, koivisto:2008xf}.  Cosmological alignment in a de Sitter background was considered in \cite{Kanno:2006ty}.   Evolution of vector field perturbations in a more general context, including the effect on primordial power spectra, was considered in \cite{Lim:2004js, Li:2007vz}.  

Here, we relax the aforementioned assumption.  We determine the dynamical evolution of the \aether\ alignment with respect to constant density hypersurfaces of flat FRW backgrounds, assuming that the \aether\ field has a negligible effect on the form of the background geometry.  
We will show that a homogeneous timelike
vector field tends to align in the presence of a homogeneous cosmological fluid such that its rest frame coincides with the rest frame of the cosmological fluid.  

Take the background spacetime to be that of a flat FRW cosmology,
\begin{equation}
ds^2 = - dt^2 + a(t)^2 (dx^2 +dy^2 + dz^2)\,.
\end{equation}
We take the equation state of the cosmological fluid to be
$p_{fluid} = w \rho_{fluid}$.   The Friedmann equation then implies 
\be a(t) = t^{2/3(1+w)} \ee
for $w \neq -1$, and 
\be a(t) = e^{Ht}\ , \quad H=\text{ constant}\ee
for $w = -1$. We assume that $m^2 / M_{P}^2$ is small, so
that the back reaction of the vector field on the FRW geometry will be small, and the
evolution of the vector field will be well approximated by its
evolution in the FRW background.

Suppose the vector field is homogeneous. This is a reasonable
assumption given that the background spacetime is homogeneous and
therefore should only affect the time evolution of the vector field. 
We may use the rotational
invariance of the FRW background to choose 
coordinates such that the $x$-axis is aligned with the spatial part of
the vector field. Then, without loss of generality, $A_0 = m \cosh
(\phi(t))$ and $A_x = m a(t) \sinh(\phi(t))$. In this case the
equations of motion reduce to,
\begin{align}
\phi''(t) + 3 H(t) \phi'(t) + \left[H^2(t) + \a H'(t)\right] \sinh(2\phi(t)) = 0,
\end{align}
where $H(t) = a'(t)/a(t)$. Expanding to first order in the angle $\phi$, for $w \neq -1$ we have
\begin{equation}
 \phi'' + \left[{2 \over (1+w)t}\right] \phi' + \left[{8 - 12\a (1+w) \over 9 (1+w)^2 t^2}\right] \phi = 0.
\end{equation}
It is a simple exercise to show that $\phi$ behaves as a damped oscillator for all $-1 < w < 1$ and $\a < {2 \over 3(1+w)}$.  For the case of a constant Hubble parameter ($w=-1$),
\begin{equation}
\phi(t) = A e^{-Ht} + B e^{-2Ht}.
\end{equation}
One can see even for large $\phi(t)$ that $|\phi(t)|$ generically decreases when $-1 < w < 1$ and $\a < {2 \over 3(1+w)}$ because, since $\sinh(\phi) = -\sinh(-\phi)$, the essential features of the full equation mirror those of the linearized equation. 

We conclude that a timelike vector field will generically tend to align to be purely timelike in the rest frame of the cosmological fluid, thereby restoring isotropy of the cosmological background. 
We do not examine the case of a spacelike \aether\ field, since that is perturbatively unstable.

\section{Extra Dimensions}

Consider now the evolution of the vector field in a background spacetime with metric
\begin{equation}
ds^2 = -dt^2 + a(t)^2 (dx^2 + dy^2 + dz^2) + dr^2.
\end{equation}
This metric is the local distance measure for a spacetime in which the infinite spatial dimensions expand as a usual flat FRW metric, for general equation of state parameter $w$ as discussed in the previous section, and a compact extra dimension with coordinate $r$ does not expand.  A scenario in which a spacelike \aether\ is aligned completely along the compact fifth extra dimension was considered in \cite{Carroll:2008pk}.

The equations of motion are once again
\begin{equation}
\label{eom5d}
(g^{\s \n} + A^\s A^\n/m^2)(\nabla_\r \nabla^\r A_\s + \a R_{\r \s}A^\r) = 0
\end{equation}
and $A_\m A^\m = -m^2$. Consider homogeneous configurations where, without loss of generality, 
\begin{equation}
A_0 = m \cosh \phi(t),~~A_x = a(t) m \sinh \phi(t) \cos \theta(t),~~A_y=A_z=0,~~\text{and}~~A_r = m \sinh \phi(t) \sin \theta(t).
\end{equation} 
The $\n = 0$ equation of motion (Eq.~\eqref{eom5d}) reads
\begin{equation}
\left[{1 \over 2}(5 - \cos 2 \theta)( H^2(1 + \a)+\a H') - 2 \a H^2 \cos^2\theta - (\theta')^2\right] \sinh 2\phi  +  6 H \phi' + 2 \phi'' = 0.
\end{equation}
When $\theta'^2 \ll H^2$, we can treat $\theta$ as being essentially constant and then the above equation determines the evolution of $\phi$. Numerical simulations indicate that $\phi$ decays to zero, whatever the value of $\theta$, if $-1 < \a < {2 \over 3(1+w)}$. One can see the decay of $\phi$ (given the bounds on $\a$) explicitly by expanding about $\phi = 0$ and $\theta = \text{constant}$ when $\phi$ is small. 

If $H$ is constant ({\emph i.e.}~the non-compact dimensions are de~Sitter-like ) and the vector field is aligned entirely along the timelike dimension and the compact dimension (so $\theta = \pi / 2$), then the equation of motion for $\phi(t)$ is
\begin{align}
\phi''(t) + 3 H \phi'(t) + {3 \over 2}(1+\a)H^2  \sinh(2\phi(t)) = 0,
\end{align}
the solution to which is
\begin{equation}
\phi(t) = A_+ e^{-\a_+ H t/2} + A_- e^{-\a_- H t/2}, 
\ee
where
\be \qquad \a_\pm = {3 }\left(1 \pm \sqrt{1- {4 \over 3}(1 + \a)}\right)\ ,
\end{equation}
when $|\phi(t)| \ll 1$.
If $1 + \a > 0$ then $\phi$ decays to zero. If $\a = -1$, $\phi$ decays to a (generically nonzero) constant, and $\phi$ can grow with time if $\a < -1$. It is interesting to see that, for the case where no perturbative modes propagate superluminally---the case where $\a = -1$---the fixed-norm vector field can evolve during a de~Sitter expansion phase so that it has a nonzero component in the compact fifth dimension while otherwise aligning so that isotropy is restored in the rest frame of the cosmological fluid. However, when the Universe enters a phase of expansion where $a(t) = t^{2 / 3(1+w)}$ and $w$ is strictly greater than $-1$ (and less than 1), then the component of the vector field in the fifth dimension will decay away.  

\section{Conclusions}

We investigated the dynamics of and limits on parameters in a theory with a fixed-norm timelike vector field whose kinetic term takes the form of a sigma model. We argued in a companion paper \cite{instabilities} that such sigma-model theories are the only \aether\ models with two-derivative kinetic terms and a fixed-norm vector field for which the Hamiltonian is bounded below.

In the presence of gravity, the action for sigma-model \aether\ is:
\begin{equation}
  S_A = \int d^4x \sqrt{-g} \left[\frac{1}{16\pi G}R- {1\over 2}( \nabla_\m A_\n)(\nabla^\m A^\n) 
  + {\alpha \over 2} R_{\m\n}A^\m A^\n 
  + {\lambda \over 2} (A_\m A^\m + m^2)\right]\,.
\end{equation}
We showed that the five massless degrees of freedom in the linearized theory will not propagate faster than light only if $\alpha = -1$, and we argued that faster-than-light degrees of freedom generically lead to instabilities on less-than-Hubble-length time scales. In the special case $\a = -1$, the vacuum expectation value $m^2$ must be less than about $10^{-4} M_p^2$, where $M_p$ is the Planck mass, in order to comply with limits on the PPN preferred frame parameter $\a_2$.  Relaxing the $\a = -1$ assumption, we summarized the strongest limits on the parameters $\{\alpha, m\}$ (from gravitational \Cerenkov\ radiation and the PPN preferred frame parameters) in Fig.~\ref{constraints fig}.  

We also showed that the \aether\ field tends to dynamically align such that it is orthogonal to constant density hypersurfaces
for the theoretically and experimentally relevant portion of the parameter space.   
The dynamics forces the rest frame of the \aether\ and that of the perfect fluid dominating the cosmological evolution to coincide.   Finally, we showed that the dynamics allows for the possibility of a non-zero spatial component in a non-expanding fifth dimension during a de~Sitter era. Even a spatial component in a non-expanding fifth dimension will decay away during non-de~Sitter eras, {\emph e.g.}, in a matter- or radiation-dominated universe. We take this as evidence that \aether\ fields with well-behaved semi-classical dynamics will not lead to any significant departure from isotropy. 

\section*{Acknowledgments} 

We are very grateful to Ted Jacobson, Alan Kostelecky and Mark Wise for helpful comments.
This research was supported in part by the U.S. Department of Energy and by the 
Gordon and Betty Moore Foundation.

\bibliography{lorentz-bib}

\begin{thebibliography}{26}
\expandafter\ifx\csname natexlab\endcsname\relax\def\natexlab#1{#1}\fi
\expandafter\ifx\csname bibnamefont\endcsname\relax
  \def\bibnamefont#1{#1}\fi
\expandafter\ifx\csname bibfnamefont\endcsname\relax
  \def\bibfnamefont#1{#1}\fi
\expandafter\ifx\csname citenamefont\endcsname\relax
  \def\citenamefont#1{#1}\fi
\expandafter\ifx\csname url\endcsname\relax
  \def\url#1{\texttt{#1}}\fi
\expandafter\ifx\csname urlprefix\endcsname\relax\def\urlprefix{URL }\fi
\providecommand{\bibinfo}[2]{#2}
\providecommand{\eprint}[2][]{\url{#2}}

\bibitem[{\citenamefont{Kostelecky and Samuel}(1989)}]{Kostelecky:1989jw}
\bibinfo{author}{\bibfnamefont{V.~A.} \bibnamefont{Kostelecky}}
  \bibnamefont{and} \bibinfo{author}{\bibfnamefont{S.}~\bibnamefont{Samuel}},
  \bibinfo{journal}{Phys. Rev.} \textbf{\bibinfo{volume}{D40}},
  \bibinfo{pages}{1886} (\bibinfo{year}{1989}).

\bibitem[{\citenamefont{Colladay and Kostelecky}(1998)}]{Colladay:1998fq}
\bibinfo{author}{\bibfnamefont{D.}~\bibnamefont{Colladay}} \bibnamefont{and}
  \bibinfo{author}{\bibfnamefont{V.~A.} \bibnamefont{Kostelecky}},
  \bibinfo{journal}{Phys. Rev.} \textbf{\bibinfo{volume}{D58}},
  \bibinfo{pages}{116002} (\bibinfo{year}{1998}), \eprint{hep-ph/9809521}.

\bibitem[{\citenamefont{Jacobson and Mattingly}(2001)}]{Jacobson:2000xp}
\bibinfo{author}{\bibfnamefont{T.}~\bibnamefont{Jacobson}} \bibnamefont{and}
  \bibinfo{author}{\bibfnamefont{D.}~\bibnamefont{Mattingly}},
  \bibinfo{journal}{Phys. Rev.} \textbf{\bibinfo{volume}{D64}},
  \bibinfo{pages}{024028} (\bibinfo{year}{2001}), \eprint{gr-qc/0007031}.

\bibitem[{\citenamefont{Eling and Jacobson}(2004)}]{Eling:2003rd}
\bibinfo{author}{\bibfnamefont{C.}~\bibnamefont{Eling}} \bibnamefont{and}
  \bibinfo{author}{\bibfnamefont{T.}~\bibnamefont{Jacobson}},
  \bibinfo{journal}{Phys. Rev.} \textbf{\bibinfo{volume}{D69}},
  \bibinfo{pages}{064005} (\bibinfo{year}{2004}), \eprint{gr-qc/0310044}.

\bibitem[{\citenamefont{Carroll and Lim}(2004)}]{Carroll:2004ai}
\bibinfo{author}{\bibfnamefont{S.~M.} \bibnamefont{Carroll}} \bibnamefont{and}
  \bibinfo{author}{\bibfnamefont{E.~A.} \bibnamefont{Lim}},
  \bibinfo{journal}{Phys. Rev.} \textbf{\bibinfo{volume}{D70}},
  \bibinfo{pages}{123525} (\bibinfo{year}{2004}), \eprint{hep-th/0407149}.

\bibitem[{\citenamefont{Jacobson and Mattingly}(2004)}]{Jacobson:2004ts}
\bibinfo{author}{\bibfnamefont{T.}~\bibnamefont{Jacobson}} \bibnamefont{and}
  \bibinfo{author}{\bibfnamefont{D.}~\bibnamefont{Mattingly}},
  \bibinfo{journal}{Phys. Rev.} \textbf{\bibinfo{volume}{D70}},
  \bibinfo{pages}{024003} (\bibinfo{year}{2004}), \eprint{gr-qc/0402005}.

\bibitem[{\citenamefont{Lim}(2005)}]{Lim:2004js}
\bibinfo{author}{\bibfnamefont{E.~A.} \bibnamefont{Lim}},
  \bibinfo{journal}{Phys. Rev.} \textbf{\bibinfo{volume}{D71}},
  \bibinfo{pages}{063504} (\bibinfo{year}{2005}), \eprint{astro-ph/0407437}.

\bibitem[{\citenamefont{Jacobson}(2008)}]{Jacobson:2008aj}
\bibinfo{author}{\bibfnamefont{T.}~\bibnamefont{Jacobson}}
  (\bibinfo{year}{2008}), \eprint{0801.1547}.

\bibitem[{\citenamefont{Carroll et~al.}(2008)\citenamefont{Carroll, Dulaney,
  Gresham, and Tam}}]{instabilities}
\bibinfo{author}{\bibfnamefont{S.~M.} \bibnamefont{Carroll}},
  \bibinfo{author}{\bibfnamefont{T.~R.} \bibnamefont{Dulaney}},
  \bibinfo{author}{\bibfnamefont{M.~I.} \bibnamefont{Gresham}},
  \bibnamefont{and} \bibinfo{author}{\bibfnamefont{H.}~\bibnamefont{Tam}}
  (\bibinfo{year}{2008}), \eprint{0812.1049}.

\bibitem[{\citenamefont{Kostelecky and Lehnert}(2001)}]{Kostelecky:2000mm}
\bibinfo{author}{\bibfnamefont{V.~A.} \bibnamefont{Kostelecky}}
  \bibnamefont{and} \bibinfo{author}{\bibfnamefont{R.}~\bibnamefont{Lehnert}},
  \bibinfo{journal}{Phys. Rev.} \textbf{\bibinfo{volume}{D63}},
  \bibinfo{pages}{065008} (\bibinfo{year}{2001}), \eprint{hep-th/0012060}.

\bibitem[{\citenamefont{Kostelecky}(2001)}]{Kostelecky:2001xz}
\bibinfo{author}{\bibfnamefont{V.~A.} \bibnamefont{Kostelecky}}
  (\bibinfo{year}{2001}), \eprint{hep-ph/0104227}.

\bibitem[{\citenamefont{Clayton}(2001)}]{Clayton:2001vy}
\bibinfo{author}{\bibfnamefont{M.~A.} \bibnamefont{Clayton}}
  (\bibinfo{year}{2001}), \eprint{gr-qc/0104103}.

\bibitem[{\citenamefont{Elliott et~al.}(2005)\citenamefont{Elliott, Moore, and
  Stoica}}]{Elliott:2005va}
\bibinfo{author}{\bibfnamefont{J.~W.} \bibnamefont{Elliott}},
  \bibinfo{author}{\bibfnamefont{G.~D.} \bibnamefont{Moore}}, \bibnamefont{and}
  \bibinfo{author}{\bibfnamefont{H.}~\bibnamefont{Stoica}},
  \bibinfo{journal}{JHEP} \textbf{\bibinfo{volume}{08}}, \bibinfo{pages}{066}
  (\bibinfo{year}{2005}), \eprint{hep-ph/0505211}.

\bibitem[{\citenamefont{Dulaney et~al.}(2008)\citenamefont{Dulaney, Gresham,
  and Wise}}]{Dulaney:2008ph}
\bibinfo{author}{\bibfnamefont{T.~R.} \bibnamefont{Dulaney}},
  \bibinfo{author}{\bibfnamefont{M.~I.} \bibnamefont{Gresham}},
  \bibnamefont{and} \bibinfo{author}{\bibfnamefont{M.~B.} \bibnamefont{Wise}},
  \bibinfo{journal}{Phys. Rev.} \textbf{\bibinfo{volume}{D77}},
  \bibinfo{pages}{083510} (\bibinfo{year}{2008}), \eprint{0801.2950}.

\bibitem[{\citenamefont{Bluhm et~al.}(2008)\citenamefont{Bluhm, Gagne, Potting,
  and Vrublevskis}}]{Bluhm:2008yt}
\bibinfo{author}{\bibfnamefont{R.}~\bibnamefont{Bluhm}},
  \bibinfo{author}{\bibfnamefont{N.~L.} \bibnamefont{Gagne}},
  \bibinfo{author}{\bibfnamefont{R.}~\bibnamefont{Potting}}, \bibnamefont{and}
  \bibinfo{author}{\bibfnamefont{A.}~\bibnamefont{Vrublevskis}},
  \bibinfo{journal}{Phys. Rev.} \textbf{\bibinfo{volume}{D77}},
  \bibinfo{pages}{125007} (\bibinfo{year}{2008}), \eprint{0802.4071}.

\bibitem[{\citenamefont{Himmetoglu et~al.}(2008)\citenamefont{Himmetoglu,
  Contaldi, and Peloso}}]{Himmetoglu:2008zp}
\bibinfo{author}{\bibfnamefont{B.}~\bibnamefont{Himmetoglu}},
  \bibinfo{author}{\bibfnamefont{C.~R.} \bibnamefont{Contaldi}},
  \bibnamefont{and} \bibinfo{author}{\bibfnamefont{M.}~\bibnamefont{Peloso}}
  (\bibinfo{year}{2008}), \eprint{0809.2779}.

\bibitem[{\citenamefont{Jimenez and Maroto}(2008)}]{Jimenez:2008sq}
\bibinfo{author}{\bibfnamefont{J.~B.} \bibnamefont{Jimenez}} \bibnamefont{and}
  \bibinfo{author}{\bibfnamefont{A.~L.} \bibnamefont{Maroto}}
  (\bibinfo{year}{2008}), \eprint{0811.0784}.

\bibitem[{\citenamefont{Mattingly}(2005)}]{Mattingly:2005re}
\bibinfo{author}{\bibfnamefont{D.}~\bibnamefont{Mattingly}},
  \bibinfo{journal}{Living Rev. Rel.} \textbf{\bibinfo{volume}{8}},
  \bibinfo{pages}{5} (\bibinfo{year}{2005}), \eprint{gr-qc/0502097}.

\bibitem[{\citenamefont{Will}(2005)}]{Will:2005va}
\bibinfo{author}{\bibfnamefont{C.~M.} \bibnamefont{Will}},
  \bibinfo{journal}{Living Rev. Rel.} \textbf{\bibinfo{volume}{9}},
  \bibinfo{pages}{3} (\bibinfo{year}{2005}), \eprint{gr-qc/0510072}.

\bibitem[{\citenamefont{Ackerman et~al.}(2007)\citenamefont{Ackerman, Carroll,
  and Wise}}]{Ackerman:2007nb}
\bibinfo{author}{\bibfnamefont{L.}~\bibnamefont{Ackerman}},
  \bibinfo{author}{\bibfnamefont{S.~M.} \bibnamefont{Carroll}},
  \bibnamefont{and} \bibinfo{author}{\bibfnamefont{M.~B.} \bibnamefont{Wise}},
  \bibinfo{journal}{Phys. Rev.} \textbf{\bibinfo{volume}{D75}},
  \bibinfo{pages}{083502} (\bibinfo{year}{2007}), \eprint{astro-ph/0701357}.

\bibitem[{\citenamefont{Dulaney and Gresham}(2008)}]{Dulaney:2008bp}
\bibinfo{author}{\bibfnamefont{T.~R.} \bibnamefont{Dulaney}} \bibnamefont{and}
  \bibinfo{author}{\bibfnamefont{M.~I.} \bibnamefont{Gresham}}
  (\bibinfo{year}{2008}), \eprint{0805.1078}.

\bibitem[{\citenamefont{Koivisto and Mota}(2008)}]{koivisto:2008xf}
\bibinfo{author}{\bibfnamefont{T.~S.} \bibnamefont{Koivisto}} \bibnamefont{and}
  \bibinfo{author}{\bibfnamefont{D.~F.} \bibnamefont{Mota}},
  \bibinfo{journal}{JCAP} \textbf{\bibinfo{volume}{0808}}, \bibinfo{pages}{021}
  (\bibinfo{year}{2008}), \eprint{0805.4229}.

\bibitem[{\citenamefont{Kanno and Soda}(2006)}]{Kanno:2006ty}
\bibinfo{author}{\bibfnamefont{S.}~\bibnamefont{Kanno}} \bibnamefont{and}
  \bibinfo{author}{\bibfnamefont{J.}~\bibnamefont{Soda}},
  \bibinfo{journal}{Phys. Rev.} \textbf{\bibinfo{volume}{D74}},
  \bibinfo{pages}{063505} (\bibinfo{year}{2006}), \eprint{hep-th/0604192}.

\bibitem[{\citenamefont{Li et~al.}(2008)\citenamefont{Li, Fonseca~Mota, and
  Barrow}}]{Li:2007vz}
\bibinfo{author}{\bibfnamefont{B.}~\bibnamefont{Li}},
  \bibinfo{author}{\bibfnamefont{D.}~\bibnamefont{Fonseca~Mota}},
  \bibnamefont{and} \bibinfo{author}{\bibfnamefont{J.~D.}
  \bibnamefont{Barrow}}, \bibinfo{journal}{Phys. Rev.}
  \textbf{\bibinfo{volume}{D77}}, \bibinfo{pages}{024032}
  (\bibinfo{year}{2008}), \eprint{0709.4581}.

\bibitem[{\citenamefont{Carroll and Tam}(2008)}]{Carroll:2008pk}
\bibinfo{author}{\bibfnamefont{S.~M.} \bibnamefont{Carroll}} \bibnamefont{and}
  \bibinfo{author}{\bibfnamefont{H.}~\bibnamefont{Tam}} (\bibinfo{year}{2008}),
  \eprint{0802.0521}.

\end{thebibliography}

\end{document}